\documentclass[aps,prb,reprint,floatfix,citeautoscript,longbibliography,superscriptaddress]{revtex4-1}
\usepackage[latin9]{inputenc}
\usepackage{amsmath}
\usepackage{amssymb}
\usepackage{graphicx}
\usepackage{mathrsfs}
\usepackage{amsfonts}
\usepackage{amsthm}
\usepackage{color,ulem}
\usepackage[colorlinks=true,citecolor=blue,linkcolor=blue,urlcolor=blue,anchorcolor=blue]{hyperref}%
\hypersetup{colorlinks=true,citecolor=blue,linkcolor=blue,urlcolor=blue}

\makeatother
\begin{document}
\title{Anomalies in the switching dynamics of C-type antiferromagnets and antiferromagnetic nanowires}
\author{H. Y. Yuan}
\affiliation{Department of Physics, Southern University of Science and
Technology, Shenzhen 518055, Guangdong, China}

\author{Man-Hong Yung}
\email[Electronic address: ]{yung@sustech.edu.cn}
\affiliation{Institute for Quantum Science and Engineering and Department
of Physics, Southern University of Science and Technology, Shenzhen 518055,
Guangdong, China}

\author{X. R. Wang}
\email[Electronic address: ]{phxwan@ust.hk}
\affiliation{Department of Physics, The Hong Kong University of
Science and Technology, Clear Water Bay, Kowloon, Hong Kong}
\affiliation{HKUST Shenzhen Research Institute, Shenzhen 518057, China}
\date{\today}
\begin{abstract}
Antiferromagnets (AFMs) are widely believed to be superior than ferromagnets in
spintronics because of their high stability due to the vanishingly small stray
field. It is thus expected that the order parameter of AFM should always align
along the easy-axis of the crystalline anisotropy. In contrast to this conventional
wisdom, we find that the AFM order parameter switches away from the easy-axis
below a critical anisotropy strength when an AFM is properly tailored into a
nano-structure. The switching time first decreases and then increases with the
damping. Above the critical anisotropy, the AFM order parameter is stable
and precesses under a microwave excitation. However, the absorption peak is
not at resonance frequency even for magnetic damping as low as 0.01.
To resolve these anomalies, we first ascertain the hidden role of dipolar
interaction that reconstructs the energy landscape of the nano-system and
propose a model of damped non-linear pendulum to explain the switching behavior.
In this framework, the second anomaly appears when an AFM is close to the boundary
between underdamped and overdamped phases, where the observed absorption lineshape
has small quality factor and thus is not reliable any longer. Our results should
be significant to extract the magnetic parameters through resonance techniques.
\end{abstract}

\maketitle
\section{Introduction}
Ferromagnets played a vital role in the early development of magnetism, as well
as the modern spintronics since late 1980s, while studies and applications
of antiferromagnets (AFMs) are quite limited for their lack of tunability, thus useless.
In the last few years, AFMs started to attract significant attention
after the discovery of electrical knob to control antiferromagnetic order in a class
of antiferromagnets with broken inversion symmetry \cite{Wadley2016,Zelezny2014}.
Various aspects, such as damping mechanism \cite{Yuan20172,Yuan20171}, spin
transfer torque \cite{Kimel2004,Duine2007,Haney2008,Xu2008}, magnetic switching
\cite{Wadley2016}, spin pumping \cite{Cheng2014}, domain wall/skyrmion
dynamics \cite{Gomonay2010,Hals2011,Helen2016,Shiino2016,Selzer2016,Jungwirth2016,
Xichao2016,Barker2016,Yuan2018a,Yuan2018b} have been extensively investigated.
One strong motivation of such intense interest in AFMs is their
abundance in nature and intriguing stability due to the vanishingly small
magnetostatic interaction (MI), which is ever doomed to be its drawback.
Accordingly, MI is neglected in most of the theoretical and numerical
studies of AFMs \cite{Hals2011,Shiino2016,Selzer2016,Helen2016}.
Nevertheless, the magnetic dipoles are there and the distribution of the dipoles
in an AFM will potentially influence the magnetic energy and thus the
magnetization dynamics, similar to the situation of electric dipoles in
dielectric materials such as liquid crystals \cite{stephen1974}.
One open question is when and how the MIs manifest themselves and influence
the magnetization dynamics.  A complete understanding of this issue may help us
in designing AFM-based devices that are truly free from the perturbation
of magnetic charges.

In this work, we take the first step to show that MI can induce a switching
of an AFM order when its crystalline anisotropy is below a critical value.
The switching occurs at an ultrafast scale and widely exists in
C-type AFMs and AFM nanowires. By analytically calculating the interaction
of magnetic charges, we find that MI produces an effective anisotropy that is
quadratic in magnetic order and thus reconstructs the energy landscape of the
system, which have observable effect on magnetization switching and spin wave
spectrum. Above the critical anisotropy, AFM resonance is
observed, but the absorption peak does not position at the true resonance
frequency when the magnetic damping is close to a critical value around 0.01.
A detailed analysis shows that the quality factor of the absorption lineshape
is significantly reduced by the
critical damping, near which the system enters the overdamped regime and
the Kittel theory based on the Lorentz lineshape fails.

This article is organized as follows. Our model, methodologies, and main findings
 are presented in Sec. II. In Sec. III, we explain the
anomalous resonance behavior near the phase boundaries and list the
typical order of critical damping for the commonly used AFMs.
Discussions and conclusions are given in Sec. IV, followed by acknowledgments.
\section{Model and results}
We first consider a two-sublattice antiferromagnetic nanowire with an
easy-axis along the longitudinal direction as shown in Fig. \ref{fig1}(a).
The magnetization dynamics is first studied by numerically solving the
Landau-Lifshitz-Gilbert (LLG) equation \cite{Vansteenkiste2014},
\begin{equation}
\frac{\partial \mathbf{S}_i}{\partial t}=-\gamma \mathbf{S}_i\times \mathbf{H}_i
+\frac{\alpha}{S} \mathbf{S}_i \times \frac{\partial \mathbf{S}_i}{\partial t},
\label{2llg}
\end{equation}
where $\mathbf{S}_i$ is the dimensionless spin vector at $i-$th site with magnitude $S$,
$\gamma$ is gyromagnetic ratio, and $\alpha$ is Gilbert damping.
$\mathbf{H}_i$ is the effective field acting on $\mathbf{S}_i$, including
antiferromagnetic exchange field between two nearest spins, crystalline
anisotropy field and stray field. The effective field can be quantitatively evaluated as
$\mathbf{H}_i=-\delta \mathcal{H}/\delta \mathbf{S}_i$.  The Hamiltonian $\mathcal{H}$ reads,

\begin{equation}
\begin{aligned}
\mathcal{H}=& J\sum_{\langle i,j \rangle} \mathbf{S}_{i}\cdot
\mathbf{S}_{j}-K\sum_i \mathbf{S}_{i,x}^2 - \frac{\mu_0\mu_s}{2S} \sum_i \mathbf{S}_{i}\cdot \mathbf{H}_{d,i},
\label{desh}
\end{aligned}
\end{equation}
where the first, second and third terms represent the exchange, crystalline
anisotropy and magnetostatic energy, respectively. $J,K,\mu_0, \mu_s$ are respectively
exchange coefficient, crystalline anisotropy coefficient, vacuum permeability and
the magnitude of local magnetic moments. $\mathbf{H}_{d,i}$ is
dipolar field acting on the spin $\mathbf{S}_{i}$. The factor 1/2 is introduced to
eliminate the duplicate calculation of magnetostatic energy.

To simulate the dynamics of the system, the parameters are taken to mimic commonly
used AFM $\mathrm{Mn_2Au}$ with $J=24$ meV \cite{Sapozhnik2018}, and $\mu_s = 3.59\mu_B$,
where $\mu_B$ is Bohr magneton.
Note that the anisotropy of $\mathrm{Mn_2Au}$ is sensitive to the magnitude
of strain \cite{Shick2010} and the magnitude of damping ($\alpha$) is still
lacking of experimental characterization, thus we treat them as free parameters.
Our main findings are: (1) The antiferromagnetic order switches spontaneously away from
the easy-axis ($x-$axis) toward the transverse direction for crystalline
anisotropy $K<1.55 ~\mu \mathrm{eV}$ (15 mT). Two typical switching events are shown
in the inset of Fig. \ref{fig1}(b). For $\alpha < \alpha_c \sim 0.01$, the
switching is accompanied by ultrafast oscillation of magnetization while
the switching is monotonic for larger dampings. (2) The switching time first
decreases and then increases with the damping and the minimum locates around the
critical damping, which separates the oscillation phase from the monotonic phase.
As a comparison, no switching happens for the ferromagnetic counterpart with
exactly the same parameters except the sign of exchange coefficient ($J$).
Next we will show that this anomalous switching of an AFM
resulting from the effect of MI and the oscillation/monotonic phase can be
understood from the underdamped and overdamped phenomena of a pendulum-like
motion of AFM order parameter.

\subsection{Theoretical formalism}
\begin{figure}
\centering
\includegraphics[width=\columnwidth]{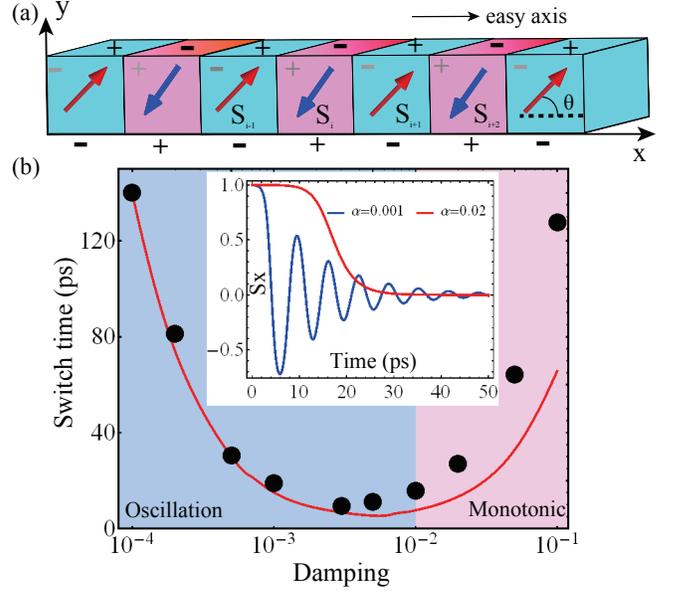}
\caption{(a) Schematic illustration of a two-sublattice antiferromagnetic nanowire.
The red and blue arrows respectively represent the magnetic moments on each
sublattice. The easy-axis of crystalline anisotropy is along the $x-$axis. The $\pm$ signs indicate
the distribution of magnetic charges on the surface (black) and inside the
volume (gray). (b) Switching time of magnetic order as a function of damping.
The red line is theoretical prediction of Eq. (\ref{dampeq}).
The light-blue and light-pink region represent the oscillating and monotonic phases, respectively.
The inset shows the two typical switching modes in the two phases for $\alpha = 0.001$ (blue line)
 and $\alpha=0.02$ (red line), respectively.}
\label{fig1}
\end{figure}

To understand the anomalous switching behavior, the key point is to properly consider
the demagnetization effect in this system. Here both the volume and surface charges
contribute to the magnetostatic field $\mathbf{H}_d$, and it can be formally
evaluated as,
\begin{equation}
\mathbf{H}_{d,i} = -M_s/S\sum_j \mathbf{N}_{ij} \cdot \mathbf{S}_j,
\end{equation}
where $M_s= \mu_s/a^3$ is the saturation magnetization, with $a$ being the
distance of two neighboring spins, $\mathbf{N}_{ij}$ is the demagnetization tensor
that depends only on the distance of two spins \cite{Newell1993}.

Suppose the system is in a N\'{e}el state with
$\mathbf{S}_i=(-1)^i S(\cos \theta \mathbf{e}_x + \sin \theta \mathbf{e}_y)$,
as shown in Fig. \ref{fig1}(a) and the longitudinal dimension $N \gg 1$, then
the total energy of the system can be calculated as,
\begin{equation}
\begin{aligned}
E (\theta)&= -(N-1)JS^2 - NKS^2\cos ^2 \theta \\
&+ Na^3K_d ( D_\parallel \cos ^2 \theta +D_\perp \sin^2 \theta ),
\end{aligned}
\end{equation}
where $K_d=\mu_0M_s^2/2$, $D_\parallel=N^{xx}_{r=0}+2\sum_{p=1}^{N/2}(-1)^pN^{xx}_{r=pa},
D_\perp = N^{yy}_{r=0} + 2\sum_{p=1}^{N/2} (-1)^p N^{yy}_{r=pa}$, $r=|i-j|a$ is
the distance between two spins. The factor $(-1)^p$ comes from the antiparallel
(parallel) alignment of two spins separating with odd (even) number of $a$,
which disappears for a ferromagnetic state. Since the magnetostatic energy of
two spins decays with their distance as $1/r^3$ \cite{jackson}, two well-separated
spins with large separation do not contribute to the energy significantly.
Here we use a cut-off distance of $r= 4a$, and analytically derive
$D_\parallel=0.5713, D_\perp=0.2144$ by evaluating the
demagnetization tensors ($N^{xx}_{r}$ and $N^{yy}_{r}$) directly.
A choice of a larger cut-off distance will not change $D_\parallel$ or
$D_\perp$ more than $1\%$.

\begin{figure}
\centering
\includegraphics[width=\columnwidth]{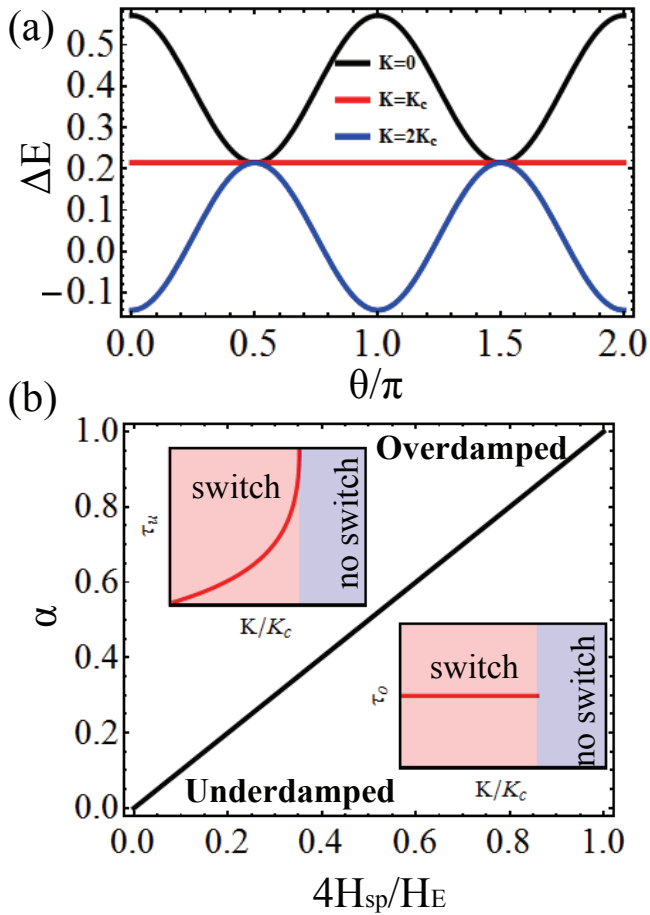}
\caption{The energy landscape of the system as a function of spin orientation
$\theta$ for $K=0$ (black line), $K_c$ (red line), $2K_c$ (blue line), respectively.
The vertical axis is scaled by $NK_d$. (b) Phase diagram of the system
in $4H_{\mathrm{sp}}/H_{E} \sim \alpha$ plane. The top-left and bottom-right insets
show the typical switching time as a function of crystalline anisotropy for
overdamped regime and underdamped regime, respectively. }
\label{fig2}
\end{figure}

Here we pay special attention to the longitudinal magnetization
states (LS, $\theta=0$) and transverse states (TS, $\theta =\pi/2$).
The energy difference of these two states can be explicitly calculated
as, $\Delta E/N= -KS^2 +K_d (D_\parallel -D_\perp)$. For an AFM with
strong crystalline anisotropy, the LS is energy preferable while the
TS becomes energy preferable when the anisotropy is very weak.
The critical anisotropy can be evaluated from $\Delta E=0$ as
$K_c= K_d (D_\parallel -D_\perp)$. Figure \ref{fig2}(a) shows the energy
landscape of the system as a function of spin orientation $\theta$ for
$K=0$ (black line), $K_c$ (red line) and $2 K_c$ (blue line), respectively.
Clearly, LS (TS) has lower energy than TS (LS) for $K>K_c$ ($K<K_c$).
Then it is expected that the antiferromagnet will spontaneously switch
from LS to TS for $K<K_c$, where the crystalline anisotropy can be reduced
by electrical means \cite{Weisheit2007,Maruyama2009,Lebeugle2009}.

To analytically describe this switching process, we recall the antiferromagnetic
dynamic equations in terms of the staggered order \cite{Yuan2018a},
\begin{equation}
\mathbf{n} \times ( \partial_{tt} \mathbf{n} +\alpha H_E \partial_t
\mathbf{n}- H_E \mathbf{h}_n ) =0 \ ,
\label{neq}
\end{equation}
where $\mathbf{n}\equiv (\mathbf{S}_{2i}- \mathbf{S}_{2i+1})/2S$ is the staggered order,
$H_E\equiv 8JS$ is homogeneous exchange field, $\mathbf{h}_n = 4(K-K_c) \mathbf{e}_x $
is the effective anisotropy field acting on the staggered order.
In spherical coordinates, the dynamic equations can be recast as,
\begin{equation}
\frac{\partial^2\psi}{\partial t^2}+2\zeta\omega_0
\frac{\partial\psi}{\partial t}+ \mathrm{sgn}(K-K_c)\omega_0^2\sin \psi =0,
\label{dampeq}
\end{equation}
where $\psi=2\theta, \zeta = \alpha H_E/(4 H_{\mathrm{sp}}), \omega_0 =
\gamma H_{\mathrm{sp}}$, $H_{\mathrm{sp}}=\sqrt {H_{\mathrm{E}}K_{\mathrm{eff}}}$
is the spin-flop field, $K_{\mathrm{eff}}=|K-K_c|$ is the effective anisotropy
coefficient that has included the contribution from MI. The sign function
$\mathrm{sgn}(x)=1$ for $x\geq0$ and $-1$ for $x<0$.
This equation is similar to the dynamics of a damped
non-linear pendulum \cite{Kim2014}. In general, the solution to Eq. (\ref{dampeq})
is an elliptic function with a complicated time dependence \cite{Kim2014}.
To have some insights on the time scale of the system, we shall solve
Eq. (\ref{dampeq}) under small amplitude approximation ($\sin \psi \sim \psi$).

According to the value of damping ratio $\zeta$, three regimes can be classified.
(i) Underdamped regime ($\zeta <1$, i.e. $\alpha < 4H_{\mathrm{sp}}/H_E$):
The solution $\psi(t)= \psi_0 e^{-\zeta \omega_0 t} \sin (\sqrt{1-\zeta^2}
\omega_0 t+\varphi_0)$. The system oscillates and decays to the equilibrium
state with a time-scale of $\Delta t= 1/ (\zeta \omega_0)$, i.e. the larger
the damping is, the faster the relaxation will be. This is consistent with
the oscillation phase in Fig. \ref{fig2}(b). (ii) Overdamped regime ($\zeta>1$,
i.e. $\alpha>4H_{\mathrm{sp}}/H_E$): $\psi(t)=\psi_0 e^{-\zeta\omega_0 t}(c_1
e^{\sqrt{\zeta^2-1} \omega_0 t} +c_2 e^{-\sqrt{\zeta^2-1} \omega_0 t})$.
Two modes $1/\tau_s =(\zeta -\sqrt{\zeta^2-1})\omega_0$ and
$1/\tau_f =(\zeta +\sqrt{\zeta^2-1})\omega_0$ compete to determine the dynamics,
while the long-time behavior of the pendulum is dominated by the slow mode $\tau_s$.
Since $\tau_s$ increases with the damping ratio, the relaxation time
becomes larger with the increase of damping. This is also consistent with the
monotonic phase in Fig. \ref{fig2}(b). (iii) Critical regime ($\zeta = 1$, i.e.
$\alpha = 4H_{\mathrm{sp}}/H_E$): $\psi(t)= \psi_0 e^{-\zeta \omega_0 t}$.
A complete phase diagram in the $4H_{\mathrm{sp}}/H_E \sim \alpha$ plane is
shown in Fig. \ref{fig2}(b). The typical overdamped and underdamped cases are
shown in the top-left and bottom-right panels, respectively.
They show distinguished anisotropy dependences.

As a comparison, the ferromagnetic counterpart of Eq. (\ref{dampeq}) reads \cite{yin2018},
\begin{equation}
\frac{\partial \theta }{\partial t} = -\alpha \gamma K_{\mathrm{eff}} \sin 2\theta
\end{equation}
which is a first-order ordinary differential equation.
This equation can be analytically solved as $-(t-t_0)/\Delta t=\ln\tan\theta$,
where $\Delta t=1/ (2\alpha \gamma K_{\mathrm{eff}})$.
Differing from antiferromagnets, the typical switching time does not
depend on the strong exchange constant $H_E$ and it usually takes a longer
time to reach the steady state because of $K_{\mathrm{eff}} \ll H_E$.

\begin{figure}
\centering
\includegraphics[width=\columnwidth]{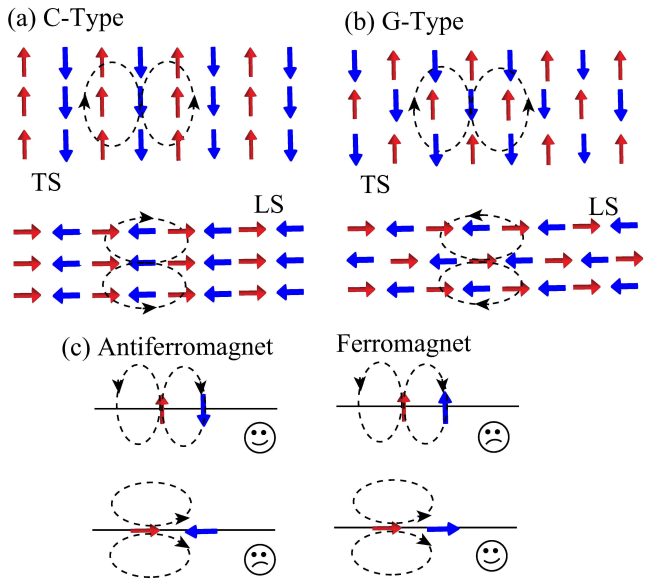}
\caption{Schematic illustration of an antiferromagnet with
 C-type ordering  (a) and G-type ordering (b) in 2D, respectively.
 The dashed lines indicate the flow of magnetostatic fields
 generated by the central spin. (c) Schematic illustration of the difference
 between two antiferromagnetically and ferromagnetically coupled spins. The smile
 and disappointed faces refer to low energy and high energy states, respectively.}
\label{fig3}
\end{figure}

\subsection{2D/3D cases}

Up till now, we focused on the switching behavior of a 1D magnetic nanowire,
but the essential physics is still valid for C-type antiferromagnet in 2D and 3D cases.
To be specific, as shown in Fig. \ref{fig3}(a), the magnetostatic
field of a particular spin (dashed line) always align parallel(antiparallel)
with the nearest spins for TS (LS) state in C-type antiferromagnet.
Hence, the TS state is energetically favorable.
For G-type antiferromagnet or checkerboard antiferromagnet, LS and TS is
energetically degenerated, which can be seen in Fig. \ref{fig3}(b).
For reference, Table \ref{tab1} lists the strength of anisotropy
coefficients induced by MI in various spin ordering of antiferromagnets,
which is calculated using the technique presented in Sec. IIA.
As the spatial dimension increases from 1D to 3D, the influence of the MI
($D_\parallel/D_\perp$) becomes more significant for C-type antiferromagnets.

\begin{table}
\begin{tabular}{|c|c|c|c|c|c|}
             \hline
             Item & 1D & 2D-CT &2D-GT& 3D-CT&3D-GT\\
             \hline
            $D_\parallel$ & 0.5713 & 0.7369 &0.4163&0.9922&0.3350\\
            \hline
            $D_\perp$ & 0.2144 & 0.0022 &0.4163&0.0028&0.3350\\
             \hline
           \end{tabular}
\caption{List of the effective anisotropy coefficients generated by magnetostatic
interaction in various spin orderings. 2D square lattice and 3D simple cubic
lattice are used to calculated these values. The symbols CT and GT are short
for C-type and G-type ordering, respectively.}
\label{tab1}
\end{table}

Before going on, we emphasize that the effective anisotropy caused by MI is
very different from ferromagnetic counterpart known as the shape anisotropy.
Use a 1D nanowire of sufficiently long length as an example, the demagnetization
factor is $D_\parallel = 0, D_\perp = 0.5$ for a ferromagnet \cite{fmdemag}, which
implies that the magnetization always tends to align in the longitudinal direction.
For an antiferromagnet, the transverse direction is preferred by MI. This difference
motivates this work that the distribution of magnetic dipoles on atomic scale
will inevitably lead to a very different energy landscape of the system.
A schematic illustration of this difference in a simple two-dipole model is given
in Fig. \ref{fig3}(c), the physics is as follows. Along a line, the head-to-tail
(ferromagnetic state) is the lowest energy state and head-to-head (anti-ferromangetic configuration)
is the highest one. On the other hand, for two dipoles in shoulder-to-shoulder,
the lower energy configuration is the antiferromagnetic arrangement, and
the ferromagnetic one is the highest one.

\subsection{Spin wave spectrum modification}

\begin{figure}
\centering
\includegraphics[width=\columnwidth]{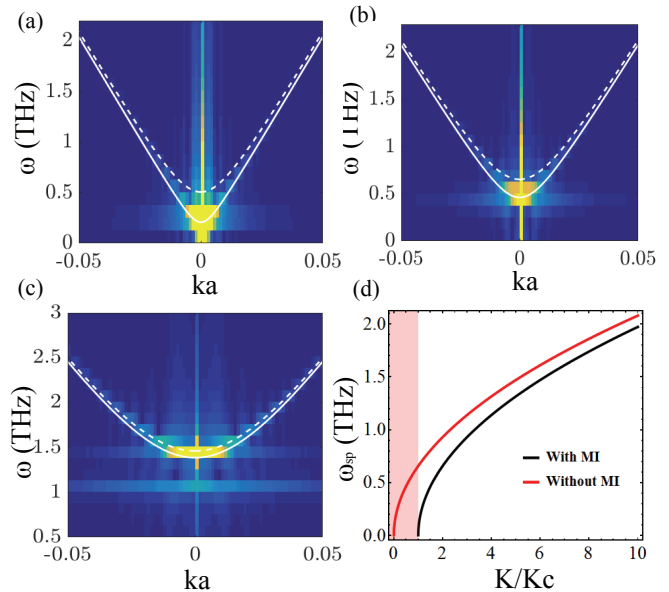}
\caption{Spin-wave spectrum of an antiferromagnetic nanowire after fully taking
account of the magnetostatic interaction. The color codes the Fourier transform
amplitude of $S_y(x,t)$. The white solid-line is the theoretical prediction with
MI while the white dashed-line is the prediction without MI.
(a) $K=1.2 K_c$, (b) $K=2 K_c$, (c) $K=10 K_c$. (d) The magnetic resonance frequency
as a function of crystalline anisotropy with (black line) and without MI (red line).
The light red region indicates the regime that spontaneous switching occurs.
$H=0$, $h_0=0.2$ T, $\omega = 10$ THz, $\alpha=10^{-4}$.}
\label{fig4}
\end{figure}

Theoretically, the spin-wave dispersion near an antiferromagnetic N\'{e}el
state is \cite{weiwei2017,yuan2017apl},
\begin{equation}
\omega = \pm \gamma H + \gamma \sqrt{H_{\mathrm{sp}}^2 + 4J^2 \sin^2 ka},
\end{equation}
where $H$ is external field, $k$ is spin wave vector. For $k=0, H = 0$,
we recover the magnetic resonance frequency $\omega_0 =\gamma \sqrt
{H_{\mathrm{E}}K_{\mathrm{eff}}}= 2\gamma \sqrt{2JK_{\mathrm{eff}}}$.
Since $K_{\mathrm{eff}} < K$ for $K>K_c$, the spin-flop field will become smaller under
the influence of MI and the spin-wave frequency tends to have a red shift.

To verify these predictions, we add a magnetic field pulse
$\mathbf{h}(t)=h_0 \mathrm{sinc}(\omega t)\mathbf{e}_y$ to excite spin-waves in an
antiferromagnetic nanowire and calculate the time dependence of
$\mathbf{S}(x,t)$ by numerically solving the LLG
equation. By taking a 2D Fourier transform of $S_y(x,t)$,
we obtain the spin-wave spectrum in $(k, \omega)$ plane as shown in Fig. \ref{fig4}(a)
($K=1.2 K_c$), \ref{fig4}(b) ($K=2 K_c$) and \ref{fig4}(c) ($K=10 K_c$).
Clearly, the dispersion can only be reproduced by including the influence of
MI (white solid-line), especially the magnetic resonance mode located at
$k=0$. As $k$ increases, the influence of anisotropy becomes small as
indicated by the merging trend of the solid and dashed lines. We also plot the comparison
of resonance frequency as a function of crystalline anisotropy in Fig. \ref{fig4}(d).
The role of magnetostatic interaction becomes most significant when $K/K_c \rightarrow 1$.

\section{Antiferromagnetic resonance}
Magnetic resonance represents large amplitude oscillation of magnetic order
when the driving frequency matches the natural frequency of the magnet.
In experiments, by measuring the position of maximum absorption and the
linewidth of resonant spectrum, one can extract the magnetic parameters such
as anisotropy and magnetic damping. In this section, we show that this common
understanding has some intrinsic problems for an antiferromagnet when the
damping is close to a critical value, which is on the order of the ratio
of spin-flop field and the exchange field ($\sim 0.01$
for $K_{\mathrm{eff}} \sim 10^{-4}H_\mathrm{E}$).

\begin{figure}
\centering
\includegraphics[width=\columnwidth]{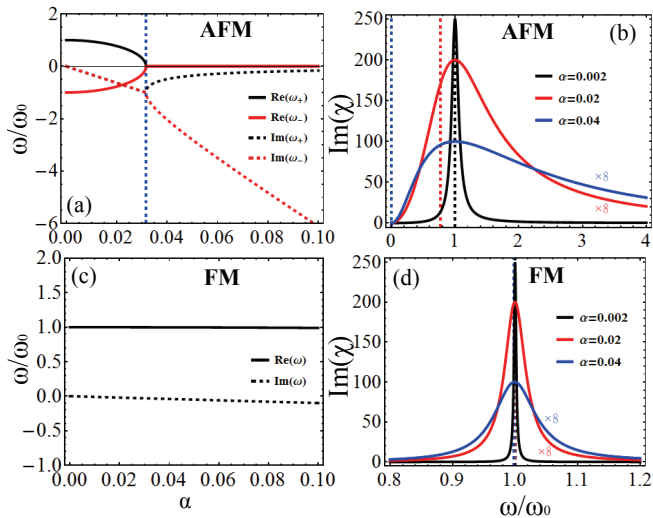}
\caption{ (a) Eigen-frequencies as a function of damping. The blue dashed line denotes
the position of critical damping $\alpha_c$. (b) Absorption spectrum as
a function of frequency when $\alpha =0.002$ (black line),~0.02 (red line),~0.04 (blue line), respectively.
The dashed lines indicate the positions of the true resonance frequency ($\omega_r$) at the corresponding damping.
(c) and (d) are the results for ferromagnets.}
\label{fig5}
\end{figure}

Let us start from the dynamic equations in terms of the two-sublattice
magnetic moments Eq. (\ref{2llg}).
Here we consider the regime $K>K_c$, By setting $\partial \mathbf{S}_i / \partial t =0$,
we find the ground state of the system is a N\'{e}el state along $x-$axis,
 as shown in Fig. \ref{fig1}(a) with $\theta =0$. Generally, the magnetic moments
will perform uniform oscillations near this ground state under the action
 of an oscillating field $\mathbf{h}=\mathbf{h}e^{-i\omega t}$,
i.e. $\mathbf{S}_{2i}=S\mathbf{e}_x + \delta \mathbf{S}_a(t),\mathbf{S}_{2i+1}=-S\mathbf{e}_x+
\delta \mathbf{S}_b(t)$. By substituting the trial solutions into Eq. (\ref{2llg})
and keeping only the terms linear in $\delta \mathbf{S}_{a,b}$, we obtain
\begin{equation}
i\frac{\partial }{\partial t} \left (\begin{array}{c}
                                \delta S_a^+ \\
                                \delta S_b^+
                              \end{array} \right )
                              =\mathbf{DH}_0 \left (\begin{array}{c}
                                \delta S_a^+ \\
                                \delta S_b^+
                              \end{array} \right )
                              + \mathbf{D}\left (\begin{array}{c}
                                -h_+\\
                                h_+
                              \end{array} \right ),
\label{ptHam}
\end{equation}
where $\delta S_a^+=\delta S_a^y+i\delta S_a^z, h_+=h_y +i h_z$.
$\mathbf{D}=\mathrm{diag}((1-i\alpha)^{-1},(1+i\alpha)^{-1})$ is dissipation matrix,
$\mathbf{H}_0$ is the effective Hamiltonian in the absence of damping,
\begin{equation}
\mathbf{H}_0 = \left ( \begin{array}{cc}
         -\Omega & -2JS \\
         2JS & \Omega
       \end{array} \right ),
\end{equation}
where $\Omega= 2JS+2K_{\mathrm{eff}}$. Then the eigen-spectrum can be
determined by solving the secular equation $\det(\omega-DH_0)=0$ as,
\begin{equation}
\omega_r=\frac{1}{1+\alpha^2} \left( -i\alpha \gamma \Omega \pm \gamma \sqrt{ H_{\mathrm{sp}}^2 -(\alpha H_E/4)^2} \right ).
\label{resf}
\end{equation}
One immediately sees that there exists a critical damping
$\alpha_c=4H_{\mathrm{sp}}/H_E$ above which the eigenfrequencies
are purely imaginary, as shown in Fig. \ref{fig5}(a). Interestingly,
this critical damping is exactly the boundary that separates the oscillation phase
(underdamped regime) from the monotonic phase (overdamped regime) discussed in Sec. IIA.

To see how the system responds to the electromagnetic wave, we can rewrite Eq.
(\ref{ptHam}) by assuming $\delta S_{a,b}(t) = \delta S_{a,b}e^{-i\omega t}$,
\begin{equation}
\left (\begin{array}{c}
                                \delta S_a^+ \\
                                \delta S_b^+
                              \end{array} \right )
                              = \left (\begin{array}{cc}
                                \chi_{aa} & \chi_{ab}\\
                                \chi_{ba} &\chi_{bb}
                              \end{array} \right )
                              \left (\begin{array}{c}
                                h_+ \\
                                h_+
                              \end{array} \right ),
\end{equation}
where
\begin{equation}
\begin{aligned}
\chi_{aa}&=\frac{\Omega - \omega -i\alpha \omega}{-\omega_0^2+(1+\alpha^2)\omega^2 + 2i\alpha \Omega \omega},\\
\chi_{bb}&=\frac{\Omega + \omega -i\alpha \omega}{-\omega_0^2+(1+\alpha^2)\omega^2 + 2i\alpha \Omega \omega},\\
\chi_{ab}&=\frac{-2JS}{-\omega_0^2+(1+\alpha^2)\omega^2 + 2i\alpha \Omega \omega},\\
\chi_{ba}&=\frac{-2JS}{-\omega_0^2+(1+\alpha^2)\omega^2 + 2i\alpha \Omega \omega}.
\end{aligned}
\end{equation}
Here we define the staggered order parameter as $\delta n = \delta S_a^+-\delta S_b^+=\chi_n h_+$, then
$\chi_n$ can be calculated as,
\begin{equation}
\chi_n = \frac{2\omega [\omega_0^2 - (1+\alpha^2)\omega^2 +2i\alpha \Omega \omega]}
{[\omega_0^2 - (1+\alpha^2)\omega^2]^2+(2\alpha \Omega \omega)^2} .
\label{chi}
\end{equation}
The imaginary part of $\chi_n$ ($\mathrm{Im}(\chi_n)$) is related to the absorption
of the system at microwave frequencies \cite{Yin2017}, which is maximal at
$\omega_m = \omega_0/(1+\alpha^2)$ \cite{omegam}. When $\alpha=0$, this peak
position is coincident with the resonance frequency predicted by
Eq. (\ref{resf}), i.e. $\omega_m = \omega_r$. Under a tiny damping, i.e. $\alpha \lll 1$,
one can reduce $\chi_n$ into the widely used Lorentz form as,
\begin{equation}
\mathrm{Im} (\chi_n) = \frac{\alpha \Omega}{(\omega-\omega_0)^2 + (\alpha \Omega)^2}.
\end{equation}
Nevertheless, as damping further increases, we notice that the peak frequency
of the lineshape ($\omega_m$) deviates from the real resonant frequency ($\omega_r$) as,
\begin{equation}
\frac{\omega_r}{\omega_m}=\sqrt{1- \left (\frac{\alpha}{\alpha_c}\right )^2}.
\end{equation}
For larger $\alpha$, the deviation of $\omega_m$ with $\omega_r$ becomes larger
and it gives completely wrong prediction of $\omega_r$ when $\alpha \sim \alpha_c$,
as shown in Fig. \ref{fig5}(b).

To resolve this anomaly, we first notice that the width of lineshape in Fig. \ref{fig5}(b) has become
comparable to the resonance frequency when $\alpha$ is close to $\alpha_c$. This suggests that
the quality factor ($Q$) of the resonance is very small and thus the lineshape is not reliable any longer.
To see this point clearly, we can solve the half-maximum width of lineshape
$\Delta \omega = 4 \alpha JS/(1+\alpha^2)$ by setting $\omega=\omega_m$ in
Eq. (\ref{chi}) and derive the $Q$ value as,
\begin{equation}
Q=\frac{\omega_m}{\Delta \omega}=\frac{\alpha_c (1+\alpha^2)}{2\alpha}\approx \frac{\alpha_c}{2\alpha}.
\end{equation}
At $\alpha=\alpha_c$, $Q =1/2$ is very bad. This effect is intrinsic for all
types of antiferromagnets, no matter the quality of the sample is high or not.
As a comparison, we can derive $Q=1/(2\alpha)$ for a ferromagnet, which does
not suffer from this problem as long as $\alpha \ll 1$, as shown in Fig. \ref{fig5}(c)
and \ref{fig5}(d).

\begin{table}
\begin{tabular}{|l|c|c|c|c|}
             \hline
             Material & $H_{E}(T)$ & $H_{\mathrm{sp}}(T)$ &$\alpha_c$& $\mathrm{Exp.}$ $\alpha$\\
             \hline
            $\mathrm{NiO} \cite{Moriyama2018}$& 524 & 39 &0.07&$5\times 10^{-4}$ \\
            $\mathrm{MnO} \cite{Sievers1963}$ & 127 & 29&0.23&$<0.02$\\
            $\mathrm{MnF_2}$ \cite{Johnson1959,Kotthaus1972} & 55.6 & 9.75 &0.18&$6\times 10^{-4}$\\
            $\alpha$-$\mathrm{Fe_2O_3} \cite{Lebrun2019}$&1040&6&0.006&NA \\
            $\mathrm{LaMnO_3} \cite{Talbayev2004}$ & 33.9 & 5.2 &0.15 & NA\\
            $\mathrm{Na_4^{3+} cluster} \cite{Nakano2013}$ &290 & 2.7 & 0.009&NA\\
            \hline
            $\mathrm{MnTe}\cite{Kriegner2017}$ & 336  & 0.5 & 0.0015 &NA\\
            \hline
            $\mathrm{Mn_2Au}\cite{Barthem2013}$ & 1300  & 5 & 0.004 &NA\\
            $\gamma$-$\mathrm{MnCu} \cite{Wiltshire1983}$ &377&13&0.034&0.78\\
             \hline
           \end{tabular}
\caption{List of the critical damping in commonly used antiferromagnets.
Note that the exchange fields have different definitions in these references,
here we only estimate the order of $\alpha_c$ as $H_{\mathrm{sp}}/H_E$
for simplicity. NA is short for ``Not Applicable" and is used when no experimental
values of dampings are found.}
\label{tab2}
\end{table}

Physically, this difference between antiferromagnets and ferromagnets comes
from their different dissipation mechanism. For an antiferromagnet, the
magnetic moments on the two sublattices must tilt away from antiparallel
orientations to launch the dissipation, while the antiferromagnetic
exchange coupling ($H_E$) tends to suppress this tendency. As the exchange
coupling  becomes large, this channel will become highly un-efficient,
therefore the resonant precession will be suppressed. For a ferromagnet,
one magnetic moment simply dissipates through the Gilbert damping.
The value of damping uniquely determines the speed of dissipation.

For reference, we summarize the typical values of critical dampings in Table \ref{tab2} that
include antiferromagnetic insulators, semiconductors and metals. They range from $10^{-3}$ to $10^{-1}$.
For most of antiferromagnetic insulators, the intrinsic damping is expected to be smaller than $10^{-3}$,
from the experience of magnetic resonance. Hence they should be well below the critical damping and
antiferromagnetic resonance is still a reliable technique to extract magnetic parameters.
For antiferromagnetic metals, the situation becomes worse since the critical damping is
considerably large compared with the real damping and the resulting lineshape may deviate significantly
from the Lorentz shape. This sets an intrinsic difficulty to analyze the resonance signal
and it is probably the reason why very few resonant experiments
are available for antiferromagnetic metals.

\section{Discussions and conclusions}

Here we would like to comment on the conventional wisdom of AFM community.
It was taken for granted that MI in AFMs is negligible without any proof.
Thus MI is neglected in most, if not all, of the analytical models, numerical
simulations and in the analysis of AFM experimental results.
Hence it is not surprising that results found here were not predicted early.
Of course, MI naturally exists in experiments, and one should be very
careful to explain the experimental data by the theory without MI effects,
especially when extracting the anisotropy coefficients.

In conclusion, we have studied MI effects on the antiferromagnetic dynamics.
Even though the total magnetic charges of an AFM as well as the resulting magnetostatic
field outside the system are vanishingly small, the local charge distribution
at atomic scale could considerably modify the system anisotropy in magnetic nanowires
as well in quasi 2D and 3D structures. By analytically evaluating the effective
dipolar anisotropy, we find that MI could even change the easy-axis of an properly
designed nano-structure. We found that the switching time first decreases
and then increases with the damping. The underdamped and overdamped phases
are thus classified, resembling the motion of a non-linear pendulum.
Near the phase boundary, the lineshape of AFM resonance becomes non-Lorentz
with very low quality factor and thus it is not reliable any more to
extract the magnetic parameters in this case.

\section{acknowledgments}
HYY acknowledge Jiang Xiao for helpful discussions.
The work is financially supported by National Natural Science Foundation
of China (NSFC) under Grant No. 61704071 and Shenzhen Fundamental
Subject Research Program under Grant No. JCYJ20180302174248595.
MHY acknowledges support by Guangdong
Innovative and Entrepreneurial Research Team Program (2016ZT06D348),
and Science, Technology and Innovation Commission of Shenzhen
Municipality (ZDSYS20170303165926217 and JCYJ20170412152620376).
XRW was supported by the NSFC Grant (No. 11774296) as well
as Hong Kong RGC Grants (Nos. 16301518, 16301619 and 16300117).


\begin{thebibliography}{99}
\bibitem{fmdemag} For a ferromagnet, we can follow a simialr approach in Sec. IIA
and derive $D_\parallel=N^{xx}_{r=0}+2\sum_{p=1}^{N/2}N^{xx}_{r=pa} \approx 0,
D_\perp = N^{yy}_{r=0} + 2\sum_{p=1}^{N/2} N^{yy}_{r=pa}\approx 0.5$.
\bibitem{omegam} One can analyticall obtain $\omega_m$ by solving the equation
$\partial \mathrm{lm} (\chi_n)/\partial \omega =0$.
\end{thebibliography}
\end{document}